\documentclass[12pt]{amsart}
\usepackage{amssymb}
\usepackage{latexsym}
\usepackage{color}
\usepackage{epic, eepic, graphicx}
\usepackage{graphicx} \DeclareGraphicsExtensions{.ps}

\newcommand{\CC}{{\mathbb C}}

\newcommand{\RR}{{\mathbb R}}

\def\t2{{\mathbb T}^2}

\newcommand{\supp}{\operatorname{supp}}

\renewcommand{\Im}{\mathop{\rm Im}\nolimits}

\newcommand{\bver}{\begin{verbatim}}

\newcommand{\defeq}{\stackrel{\rm{def}}{=}}
\def\hto0{\xrightarrow{h\to 0}}

\theoremstyle{plain}

\theoremstyle{definition}

\newcommand{\bequ}{\begin{equation}}

\def\bbbone{{\mathchoice {1\mskip-4mu {\rm{l}}} {1\mskip-4mu {\rm{l}}}
{ 1\mskip-4.5mu {\rm{l}}} { 1\mskip-5mu {\rm{l}}}}}

\def\squarebox#1{\hbox to #1{\hfill\vbox to #1{\vfill}}}

\usepackage{epic, eepic}

\title[Symmetry of bound and antibound states]
{Symmetry of bound and antibound states in the semiclassical limit} 

\author[D. Bindel]{David Bindel}
\author[M. Zworski]{Maciej Zworski}
\address{Department of Mathematics,
Courant Institute of Mathematical Sciences\\
New York University, New York, NY 10012}
\email{dbindel@cims.nyu.edu}
\address{Mathematics Department, University of California \\
Evans Hall, Berkeley, CA 94720, USA}
\email{zworski@math.berkeley.edu}

\begin{document}

\maketitle

\section{Introduction and statement of the theorem}

The simplest model of scattering/quantum resonances comes from 
considering compactly supported potentials on the real line, 
\begin{equation}
  \label{eq:V} 
   V(x)  \in \RR, \ \  
  |V(x)| \leq C,  \ \ 
  V(x) = 0 \text{ for }|x| > L \,,
\end{equation}
and the corresponding Schr\"odinger operators,
\begin{equation}
  \label{eq:Hv}
  H_V \defeq - \partial_x^2 + V(x),
\end{equation} 
on $ \RR $, or on $ [0, \infty ) $ with Dirichlet or Neumann boundary 
conditions.

The {\em resonances} or {\em scattering poles} of $ H_V $ are defined
as the poles of the meromorphic continuation of the resolvent,
$  R_V(\lambda) =
(H_V - \lambda^2)^{-1} $,
from $   \Im \lambda > 0$, to $ \CC $. Except for the poles at $ \lambda
$ for which $ \lambda^2 $ are eigenvalues of $ H_V $, $ R_V ( \lambda ) $
is bounded on $ L^2 $ for $ \Im \lambda > 0 $. Its Schwartz kernel, that
is the Green function, continues meromorphically across the continuous
spectrum corresponding to $ \RR $. Its poles are the resonances of 
$ H_V $.

An illustration based on the numerical codes of \cite{BiZw} is given 
in Fig.\ref{f:class}. The poles on the positive imaginary axis correspond
to the bound states of $ H_V $, and the poles on the negative are called
{\em antibound states}. Note that they appear to be exactly symmetric
with the bound states.
In this note we prove a simple theorem inspired by numerical experiments
using \cite{BiZw}:

\medskip
\noindent
{\bf Theorem.}{\em Consider the Dirichlet (or Neumann) boundary condition on $ [0 , \infty ) $
and a compactly
supported piecewise continuous 
potential $ V_0 $, $\supp V_0  \subset [ 0 , A )  $.
Let $ V_1 > 0 $, $ B > A $, and put 
\[ V ( x ) = V_0 (x ) + \bbbone_{[A,B]} ( x ) V_1\,. \]
Then 
the bound and antibound states of $ H_{q^2 V } $
with moduli greater
than some fixed $ k_0 > 0 $ are symmetric modulo
errors of size $ e^{-c q} $,  $ q \rightarrow \infty $.}

\medskip

Equivalently we can consider the semiclassical problem 
\[ ( - ( h \partial_x)^2 + V ( x ) ) u ( x ) = z(h) u ( x ) \,,\]
for which the conclusion of the theorem says that bound and antibound
states with moduli greater than $ h k_0 $ are symmetric modulo 
exponentially small errors, $ \exp ( -c/h ) $, as $ h \rightarrow 0 $.

\begin{figure}[ht]
\begin{center}
$$\hspace{-0.75cm}\includegraphics[width=17cm]{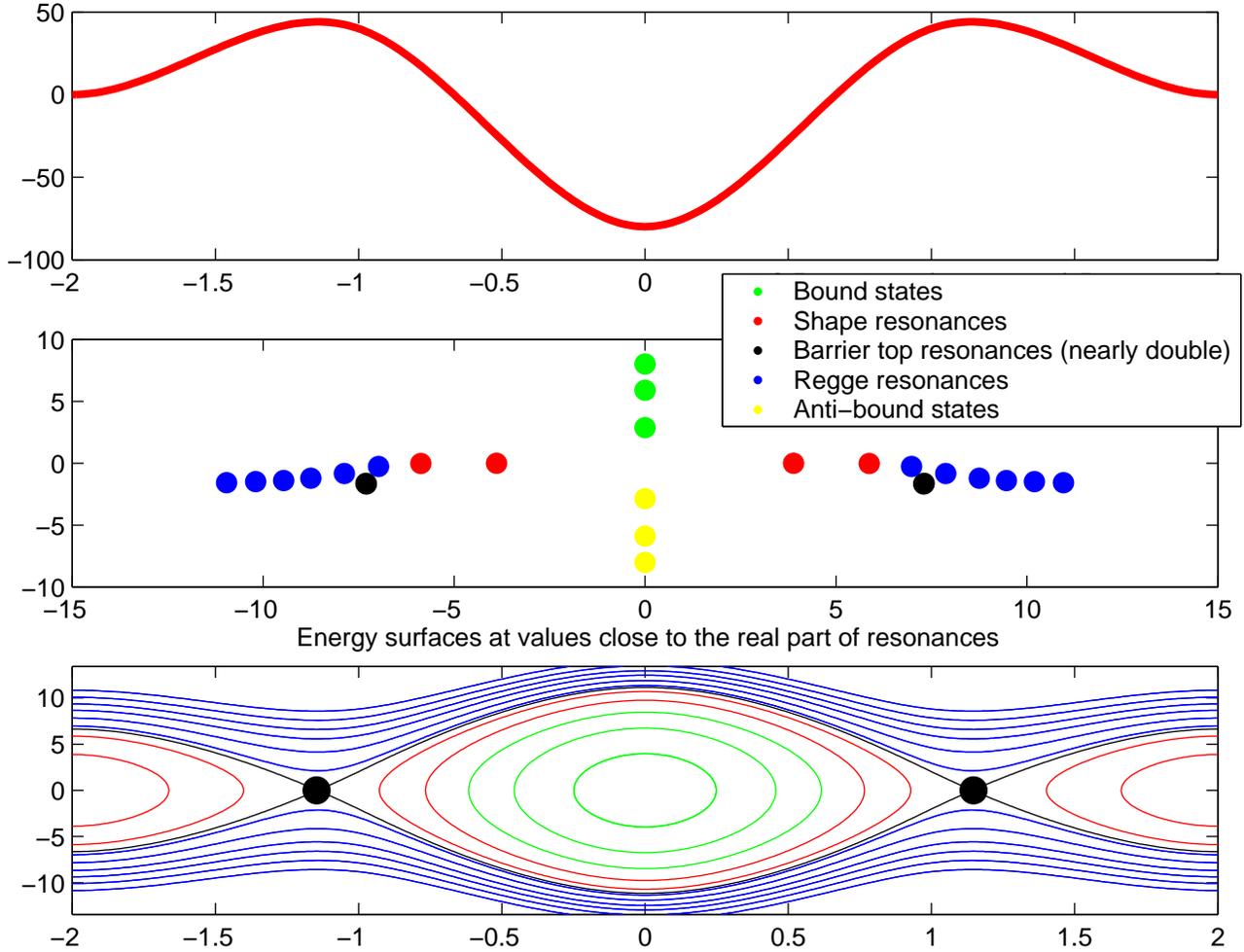}$$
\end{center}
\caption{The (color coded) correspondence between classical
dynamics and the distribution of resonances. The $ C^1 $
potential and its resonances are obtained using
{\tt splinepot(40*[0,1,-2,1,0],[-2,-1,0,1,2])} from \cite{BiZw}.}
\label{f:class}
\end{figure}

We think of $   \bbbone_{[A,B]} ( x ) V_1 $ as a barrier separating 
the potential in the interaction region, $ V_0 ( x)  $, from 
infinity. The same results hold on the line but the proof becomes slightly 
more cumbersome to write. In Fig.\ref{f:exV} we show an example of 
potentials $ V_0$, $ W =  V_1 \bbbone_{[A,B]}  $, and $ V = V_0 + W $. 
It is quite possible that using more sophisticated methods -- see
for instance \cite{FuRa} and \cite{ThRa} -- more general barriers
can be considered. Our goal here was to present a simple new 
result discovered by a numerical observation.
It is easy to see (\cite{Si} and \S \ref{pr} below) that for the 
problem on the half line the bound and antibound states are {\em never}
exactly symmetric. Yet, in a presence of a mild barrier, they are
symmetric within numerical accuracy of a computation: the exponential 
convergence is indeed very rapid. This is illustrated in Fig.\ref{f:bab}:
we plots of imaginary parts of 
bounds states and negatives of the imaginary parts of antibound states
for $ q^2 V ( x ) $ as a function of $ q^2 $. The difference
between the two pictures is striking. As expected the bound states
are not much affected by the presence of  $ W $ but the antibound
states change dramatically and in the presence of a barrier 
become nearly symmetric (this is a curious pseudospectral effect).
The high lying states of $ V_0 $ also
exhibit the symmetry -- experiments show that it is always so, even 
when there is no barrier,  and it 
improves for more regular potentials.

The study of resonances/scattering poles in one dimension 
has a long tradition 
going back to origins of quantum mechanics, see for instance \cite{LL}.
Perhaps the first study of their distribution 
was conducted by Regge \cite{Reg}.
For mathematical results in one dimension see
\cite{AAD},\cite{Fr},\cite{Hi},\cite{Kor},\cite{Ned},\cite{Si},\cite{Zw},
and many other articles. Concerning antibound states, Hitrik \cite{Hi}
showed (using a Ricatti equation approach which we also find useful in 
\S \ref{pr}) that for positive compactly supported potentials, there
are no antibound states in the semiclassical limit. That of course 
corresponds to our result since there are no bound states either. 
Simon \cite{Si} showed that for a half line problem existence of 
$ n $ bound states implies the existence of $ n - 1$ antibound states. 
Since the set of resonances of an even potential is the union of 
Dirichlet and Neumann resonances of the half line problem, this means
that having $ n $ bounds states implies the existence of $ 2 n - 2 $ 
antibound states. As can be checked using \cite{BiZw} this is often
optimal for negative potentials but never for potentials with a barrier.

Our note is organized as follows: in \S \ref{pr} we give the elementary 
proof of the theorem and in \S \ref{num} we describe the ideas behind
the computation of resonances in one dimensions. The {\tt MATLAB}
codes based on 
that section are available at \cite{BiZw}.

\medskip

\noindent
{\sc Acknowledgments.} 
The work of the second author was supported in part by a
National Science Foundation grant DMS-0200732.

\section{Proof of the theorem}
\label{pr}

\begin{figure}[ht]
  \begin{center}
    \includegraphics[width=16cm]{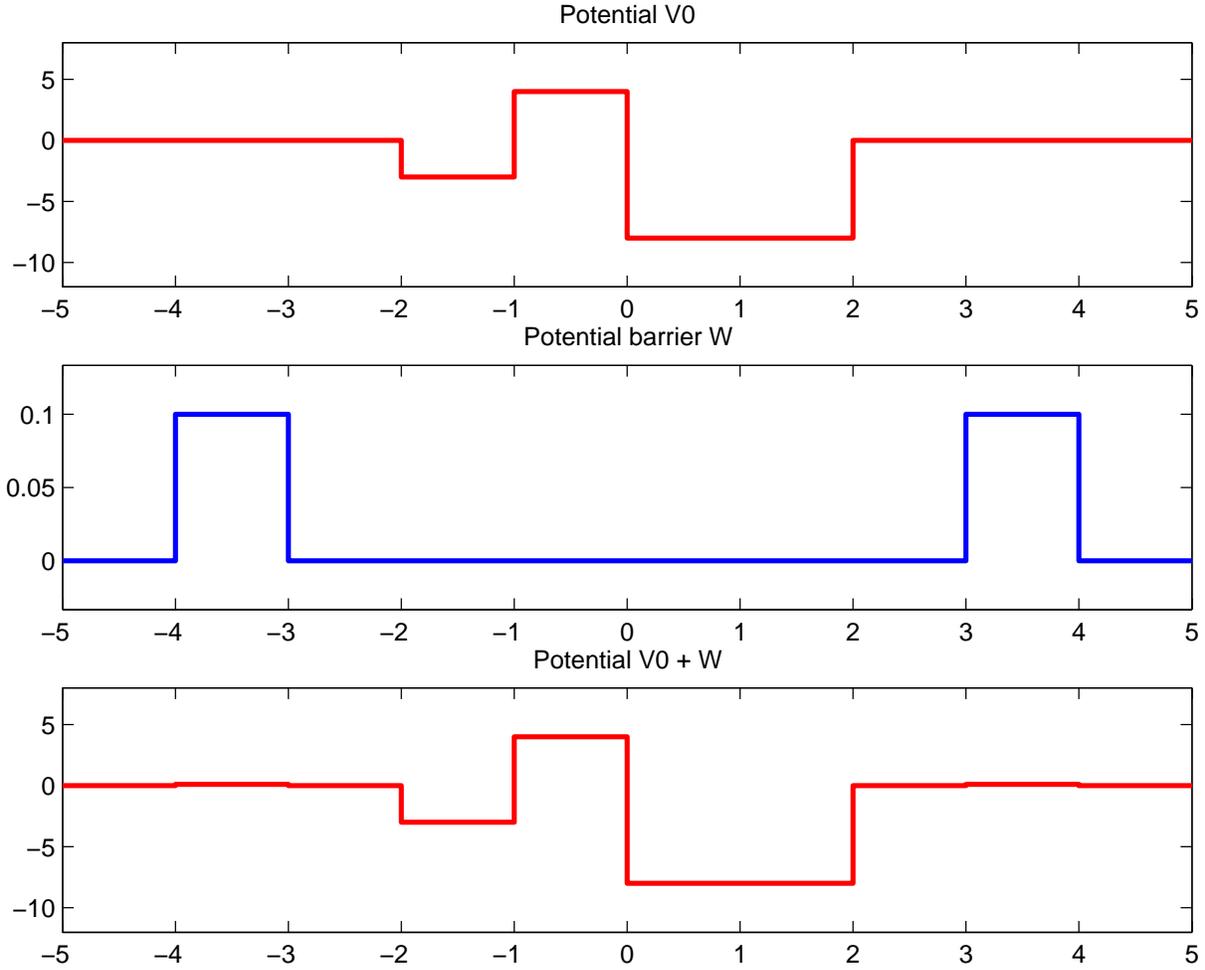}
  \end{center}
  \caption{Potentials with bounds and antibound states shown in 
Fig.\ref{f:bab}.}
  \label{f:exV} 
\end{figure}

\begin{figure}[ht]
  \begin{center}
  \includegraphics[width=16cm]{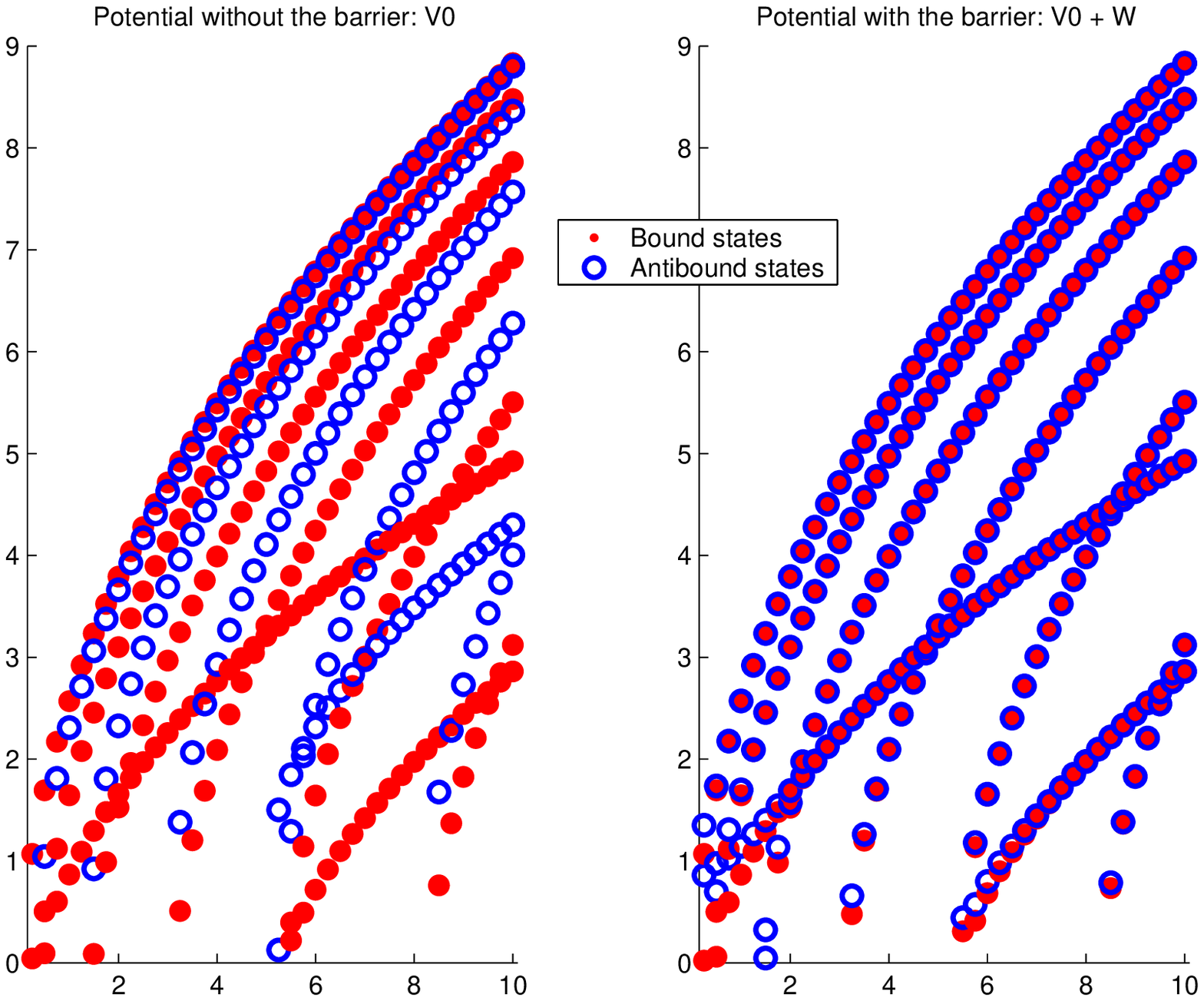}
  \end{center}
  \caption{Imaginary parts, $ \lambda $, of bounds states (corresponding to 
eigenvalues at $ - \lambda^2 $), and negatives of the 
imaginary parts of antibound states of $ q^2 V_0 $, and 
of $ q^2 ( V_0 + W ) $, as functions of $ q^2$.}
  \label{f:bab} 
\end{figure}

We will prove the theorem for $ A = 1 $ and $ B = 2 $ -- the general
case is identical. 
We consider transfer operators for solutions of $ H_{q^2V} + k^2 $:
\[ M_0 ( k )  : [ u ( 0 ) , u'( 0) ] \mapsto [ u(1) , u'( 1) ] \]
which depends only on $ V_0 $, $ q $, and $ k $, and
\[ M_1 ( k )  : [ u ( 1 ) , u' ( 1 ) ] \mapsto [ u ( 2 ) , u' ( 2) ] ]\,, \]
which is completely explicit since we know $ V_1 $:
\begin{equation}
\label{eq:m1}
 M_1 ( k ) = \frac{ 1}{ k_1 } \left( \begin{array}{ll} k_1 \cosh k_1 
& \ \sinh k_1 \\ k_1^2 \sinh k_1 & k_1 \cosh k_1 \end{array} \right) \,, 
\ \  k_1 = \sqrt { k^2 + q^2 V_1 } > k + q/C \,.
\end{equation}

As in the code described in \S \ref{num}, $ i k $, $ k > 0 $, is a bound state if and 
only if 
\[ M_1 ( k ) \circ M_0 ( k ) \, [ 0, 1 ] = [  \alpha, -k \alpha ] \,, \]
for some $ \alpha $, and $ -i k $ is an antibound state if and only if 
\[ M_1 ( k ) \circ M_0 ( k ) \, [ 0, 1 ] = [ \beta , k \beta ] \,, \]
for some $ \beta $ (note that same $ k $ will never do for both,
so they are {\em never} exactly symmetric).

The
conditions for a bound ($-$) and an antibound state ($+$) then
become (note that the left hand side depends on $ k $ and $ q $ only, and 
not on $ \pm $):
\begin{equation}
\label{eq:v}  \frac{ u'( 1 ) }{ u ( 1 ) } = - k_1
\frac{ 1 - \beta_\pm ( k) \exp ( - 2 k_1 ) }{ 1 + \beta_\pm ( k  )
\exp( - 2 k_1)}\,, \ \ \beta_\pm( k ) = \frac{ 1 \pm k/k_1} { 1 \mp k/k_1 } \,. \end{equation}

In fact, putting 
$ v ( k ) \defeq u' ( 1 ) / u ( 1 ) $,
we use \eqref{eq:m1} to 
obtain the following equations for bound/antibound states:
\[ \pm \frac{k}{k_1} ( k_1 ( 1 + e^{-2k_1}) + v ( k ) ( 1 - e^{-2k_1 } ) 
) = k_1 ( 1 - e^{-2 k_1 } ) + v ( k ) ( 1 + e^{-2 k_1} ) \,, \]
or 
\[ k_1 \left( \pm \frac{k}{k_1}  ( 1 + e^{-2k_1}) - ( 1 - e^{-2k_1 } )\right)
 = v(k ) \left(  1 + e^{-2 k_1 }  \mp \frac{k}{k_1}  
( 1 - e^{-2 k_1} ) \right) \,, \]
from which \eqref{eq:v} follows.

The behaviour of $ v ( k ) $  as $ k $ varies is well known:
it is monotonic between $ - \infty $ and $ \infty $ where $ \infty $
correspond to $ k^2 $ which are Dirichlet eigenvalues of $ H_{q^2V_0 } $
on $ [ 0 , 1 ] $. But the equations for $ \pm $ are the same
up to exponentially small errors!

More precisely, suppose that
\[ u'' = (k^2 + V ( x ) ) u \,, \ \ u ( 0 ) = 0 ,\ \  u'( 0 ) = 1\,, \]
$ u = u ( x , k ) $.
Consider
\[ v ( x , k ) \defeq  u'( x , k ) / u( x , k ) \,.\]
Then $ v $ satisfies the Ricatti equation
$$ v' = k^2 + V - v^2 \,. $$
Denote differentiation with respect to $ k $ by $ \dot v $.
We get the following equation for $ \dot v$:
$$ \dot v' = 2 k - 2 v \dot v \,, $$ 
which we can solve by the method of integrating factor.
Noting that $ u ( x )^2 \dot v ( x ) |_{ x = 0 } = 0 $
(from the boundary conditions) we get
\[ \dot v ( x , k ) = \frac{ 2 k }{ u(x,k)^2} \int_0^x u ( y , k)^2 dy \]
and in particular we get an expression for the derivative of the
Dirichlet to Neumann map:
\[ \dot v ( 1 , k ) = \frac{ 2 k }{ u(1,k)^2} \int_0^1 u ( y , k)^2 dy \]
which we can estimate from below as follows.

Since we assumed that $ \supp V \subset [ 0, 1) $, for some $ \epsilon > 0 $
we have
\[  u ( x, k ) = A e^{xk } + B e^{-xk} \,, \ \ 1 - \epsilon < x \leq 1 \]
Then
\[ v (  1 , k ) = k \frac {\alpha- 1} { \alpha+ 1} \,, \ \ 
\alpha\defeq \frac A B e^{ 2 k } \,, \]
and using the same notation,
\begin{equation}
\label{eq:dotv}
 \begin{split} \dot v ( 1 , k ) & \geq \frac{ 2 k } { u ( 1, k )^2}
\int_{1-\epsilon}^1 u ( x , k )^2 dx \\
& = \frac{ \alpha^2 ( 1 - e^{-2k\epsilon} ) +  e^{ 2k \epsilon} 
( 1 - e^{-2k \epsilon } ) + 4 \epsilon k \alpha}
{ ( \alpha+ 1 ) ^2 } \\
& \geq  \frac{ ( 1 - \delta ) ( \alpha+ 1)^2 + \delta e^{2 k \epsilon}/ C }{ 
( \alpha+1 )^2 }  
\\ 
& \geq 1 - \delta 
\,, \  \ k \geq k_0 ( \epsilon, \delta ) \,, 
\end{split} 
\end{equation}
for any $ \delta > 0 $.

We recall that the condition \eqref{eq:v} for being a bound ($-$) or
an antibound ($+$) state was
\[  v ( 1 , k_\pm ) = ( k_\pm^2 + q^2 V_1 )^{\frac12} ( 1 + g_{\pm} ( k_\pm , q ) ) \]
where $ g_\pm = {\mathcal O} ( e^{ - c q  } ) $.
Put 
\[ F ( k ) \defeq \frac{ v ( 1 , k ) }{ (k^2 + q^2 V_1 )^{\frac12} } - 1\,,
\]
so that $ F ( k_\pm ) = g_\pm ( k_\pm , q ) =  {\mathcal O} ( e^{ - c q  } ) $. 
Once we show that  $ \dot F( k ) \neq 0$, we will know that
the roots of $F$ are stable, and by standard theory,
small perturbations to the equation lead only to small perturbations
to the roots. 

More precisely, we use \eqref{eq:dotv} to estimate
\[ \begin{split} \dot F ( k ) & = 
 \frac{ \dot v ( 1 , k ) ( k^2 + q^2 V_1)
- k v ( 1 , k ) } { ( k^2 + q^2 V_1 )^{\frac{3}{2}}} 
\\ 
& \geq 
\frac{  ( 1- \delta) ( \alpha + 1)^2  
( k^2 + q^2 V_1)
 - k^2 ( \alpha^2 - 1) } { ( k^2 + q^2 V_1 )^{\frac32} ( \alpha
 + 1 )^2 } \\
& \geq  \frac{\delta} {k} \,, \ \ k > k_0 ( \epsilon , \delta ) \,,
\end{split}
\]
provided that $ \delta $ is taken small enough depending on $ V_1 > 0 $. 
Hence, by the mean 
value theorem there exists some $ 0 < s < 1 $, such that 
\[ \begin{split} | k_+ - k_- | & 
= \frac{| F(k_+ ) - F( k_-)| }{ | \dot F ( ( 1 -
s) k_+ + s  k_- )| } \\ & \leq C { e^{-2c q} } {(  ( 1 -
s) k_+ + s  k_-  ) } /\delta \\ & \leq 
e^{ - cq } \,, \ \ k_\pm > k_0   \,, \ \ q > q_0 \,. \end{split} \]
Note that we used the fact that $ k_\pm $ are necessarily bounded 
by $ C q $.

Replacing the explicit solutions by WKB approximations might
give a more general result.

\section{Numerical computation of resonances in one dimension}
\label{num}

In this section we describe the ideas behind the codes, 
{\tt squarepot.m} and {\tt splinepot.m}, used to produce Fig.\ref{f:class}
and \ref{f:bab}. These {\tt MATLAB} codes are available at \cite{BiZw}.

If the support of $V$ is contained in a compact interval $[-L,L]$,
we can compute both resonance solutions and ordinary eigenvalues of the
Schr\"odinger problem, $ (H_V - \lambda^2) u = 0 $,
 by writing appropriate boundary
conditions at $\pm L$:
\begin{equation}
\begin{array}{ll}
  (H_V - \lambda^2) u = 0        & \mbox{ for } x \in (-L,L), \\
  (\partial_x + i \lambda) u = 0 & \mbox{ at } x =  L, \\
  (\partial_x - i \lambda) u = 0 & \mbox{ at } x = -L.
\end{array}
\label{eq:bdd_eig1}
\end{equation}
In terms of $\lambda$, this is a \emph{quadratic eigenvalue problem}.
We can introduce a new variable $\psi = \lambda u$ to convert this problem
to a linear eigenvalue problem in two fields:
\begin{equation}
\begin{array}{ll}
  H_V u - \lambda \psi = 0       & \mbox{ for } x \in (-L,L), \\
  \lambda u - \psi = 0           & \mbox{ for } x \in [-L,L], \\
  (\partial_x - i \lambda) u = 0 & \mbox{ at } x =  L, \\
  (\partial_x + i \lambda) u = 0 & \mbox{ at } x = -L.
\end{array}
\label{eq:bdd_eig2}
\end{equation}
We now discretize the boundary and domain operators to get a
finite-dimensional generalized eigenvalue problem.
For small discretizations with up to a few hundred unknowns, we can
solve this generalized eigenvalue problem using MATLAB's \texttt{eig}
command, which uses the dense eigensolvers in LAPACK~\cite{LAPACK}.  For
larger discretizations, we use MATLAB's \texttt{eigs} to call ARPACK,
a standard Arnoldi-based iterative eigensolver~\cite{ARPACK}.

For the calculations shown in this note, we used a high-order
pseudospectral collocation method to discretize the
operators~\cite{Trefethen}, \cite{Boyd}.  We partition the support
interval $[-L,L]$ into subintervals, and approximate $u$ by a
high-order polynomial on each subinterval.  At the Chebyshev points on
the interior of each subinterval, we insist that the domain
differential equations be satisfied exactly, while at the junctions
between neighboring intervals, we insist that the solution $u$ and the
first derivative $\partial_x u$ must both be continuous.  Assuming
that the potential is smooth except possibly at the endpoints of the
subintervals, the collocation scheme we use is {\em spectrally
accurate}; that is, the error asymptotically decreases faster than
any algebraic function of the order of the collocation scheme.  As a
simple check on the accuracy of the computed eigenvalues of
(\ref{eq:bdd_eig2}), we increase the order of the method by 50\%,
recompute the eigenvalues, and compare the results obtained from the
coarser and the finer discretization.

We can write the analogue of (\ref{eq:bdd_eig1}) in higher dimensions,
with a {\em Dirichlet-to-Neumann (DtN) map} -- or some approximation to a
DtN map -- in place of the boundary conditions at $\pm L$.  In more
than one space dimension, this boundary map ceases to be a linear
function of $\lambda$, and so we cannot easily convert the problem
into a linear eigenvalue problem.  Researchers are studying these more
complicated \emph{nonlinear eigenvalue problems} for a variety of
engineering problems~\cite{Bai}.  Many of these problems involve
resonances in models of elastic, acoustic, or electromagnetic
resonators with radiation losses.

For comparison,  we  will also discuss other methods for computing 
resonances. They are essential for effective codes for
higher dimensional problems for which analogues of \eqref{eq:bdd_eig1} 
are unavailable or become more complicated.

Often, resonances are computed by changing the equation so that it is no
longer posed on all of $\RR$, but instead is posed on some interval $(-M,M)$
with homogeneous Dirichlet or Neumann boundary conditions.  For example,
if the support of $V$ lies strictly within the interval $(-L,L)$, we might
add a {\em complex absorbing potential} outside of $(-L,L)$, or we might
scale the coordinate system into the complex plane by the method of
{\em perfectly matched layers}\footnote{See \cite{Da} for a comparison of that 
method with the complex 
scaling method described, for instance, in \cite{TZ}}.  The change to the equation should be
designed so that the modified equation mimics the behaviour of the original
problem in the range $(-L,L)$.

To be more concrete, suppose that we modify the equation on the interval
$(L,M)$ so that we still have a nonsingular, second-order, ordinary
differential equation in $x$ whose coefficients depend on $\lambda$.
Now we specify two linearly independent solutions $\gamma_+(x,\lambda)$ 
and $\gamma_-(x,\lambda)$ on $(L,M)$ which satisfy the modified domain
equation together with the initial conditions
\begin{equation}
  \begin{array}{l}
  \gamma_+(L,\lambda) = 1, \ \ \partial_x \gamma_+(L, \lambda) = i \lambda \\
  \gamma_-(L,\lambda) = 1, \ \ \partial_x \gamma_-(L, \lambda) = -i \lambda.
  \end{array}
  \label{eq:abc_ics}
\end{equation}
These initial conditions are consistent with the conditions for
outgoing and incoming waves on $(L-\epsilon, L)$.  Now suppose
that $\gamma(x, \lambda)$ satisfies the differential equation on
$(L,M)$, and also the boundary condition $\gamma(M,\lambda) = 0$.
Then
\begin{equation}
  \gamma(x,\lambda) = 
  c \left( \gamma_+(x, \lambda) + \rho \gamma_-(x, \lambda) \right)
  \label{eq:abc_gamma}
\end{equation}
where $c$ is an arbitrary constant and
\[
  \rho(\lambda) \defeq 
    -\frac{\gamma_+(M, \lambda)}
          {\gamma_-(M, \lambda)}
\]
is a constant whose amplitude reflects how well the equation on $(L,M)$
serves to absorb outgoing waves.  We can therefore convert the condition
at $x = M$ to a condition at $x = L$.  Subsituting (\ref{eq:abc_ics})
into (\ref{eq:abc_gamma}), we have
\[
  \partial_x \gamma(L) - 
  i \lambda 
  \left( \frac{1-\rho(\lambda)}{1+\rho(\lambda)} \right) \gamma(L) = 0,
\]
which, for regions of the complex plane where $|\rho(\lambda)|$ is small, can
be treated as a perturbation of the exact outgoing wave condition at $L$.

In summary, by changing the Schr\"odinger equation outside the 
interval $(-L,L)$,
imposing homogeneous Dirichlet boundary conditions at $\pm M$, 
and then transporting the conditions at $\pm M$ to conditions at $\pm L$,
we arrive at the equations
\begin{equation}
\begin{array}{ll}
  (H_V - \lambda^2) \hat{u} = 0        & \mbox{ for } x \in (-L,L), \\
  (\partial_x + i \hat{\lambda}) \hat{u} = 0 & \mbox{ at } x =  L, \\
  (\partial_x - i \hat{\lambda}) \hat{u} = 0 & \mbox{ at } x = -L.
\end{array}
\label{eq:abc_eig1}
\end{equation}
where
\[
  \hat{\lambda} \defeq 
    \lambda \left( \frac{1-\rho(\lambda)}{1+\rho(\lambda)} \right).
\]
For values of $\lambda$ where $|\rho(\lambda)| \ll 1$, (\ref{eq:bdd_eig1})
and (\ref{eq:abc_eig1}) may be treated each as a perturbation of the other.
We note that  $\rho(\lambda)$ and $\beta_{\pm}(k) \exp(-2k_1)$ of 
\eqref{eq:v} play  similar r\^oles in the two situations. However, 
the smallness of $ \rho ( \lambda ) $ is achieved through ellipticity due to 
the complex deformation, and the smallness of  $\beta_{\pm}(k) \exp(-2k_1)$
is due to the presence of a real barrier, $ V_1 \bbbone_{[0,1]} $.

The relation between outgoing wave boundary conditions and wave behaviour
at the boundary of a bounded absorber is useful for applications and
experiments as well as for calculations.  Experiments to observe acoustic
(or electromagnetic) resonances and scattering are generally conducted in 
{\em anechoic chambers}, which are lined with baffles of sound-absorbing
material.  These baffles prevent incoming reflected waves from interfering
with the experiment.  Just as one can mimic the ``radiation-only'' property
of an infinite domain with a finite absorber, models set in infinite domains
are often approximations of models over a large finite domain in which
the medium through which waves propogate is slightly dissipative.


\begin{thebibliography}{XX}

\bibitem{AAD} A.A.~Abramov, A.~Aslanyan, and E.B.~Davies,
{\em Bounds on complex eigenvalues and resonances}, 
J. Phys. A {\bf 34}(2001), 57--72.

\bibitem{LAPACK}
E. Anderson, Z. Bai, C. Bischof, S. Blackford, J. Demmel, J. Dongarra,
J. Du Croz, A. Greenbaum, S. Hammarling, A. McKenney, and D. Sorensen.
\emph{LAPACK Users' Guide.} SIAM, 1999.

\bibitem{Bai}
Z. Bai and C. Yang, 
 {\emph{From Self-Consistency to SOAR: Solving Large-Scale
Nonlinear Eigenvalue Problems,}} SIAM News {\bf 39}(3)(2006).

\bibitem{BiZw} D. Bindel and M. Zworski,
Resonances in one dimensional: theory and computation (including
{\tt MATLAB} codes), {\tt www.cims.nyu/$\sim$dbindel/resonant1D}


\bibitem{Boyd}
J.P. Boyd, 
 {\emph{Chebyshev and Fourier Spectral Methods.}}
Dover, 2001.

\bibitem{Briet} P. Briet, J.-M. Combes, and P. Duclos,
{\em On the location of resonances in the semi-classical limit II,}
Comm. Math. Phys. {\bf 12}(1987), 201-222.

\bibitem{Da} K. Datchev, 
 {Computing resonances by generalized complex scaling.} \\
{\tt http://math.berkeley.edu/$\sim$datchev/resonance3.ps}




\bibitem{Fr} R. Froese, {\em Asymptotic distribution of resonances in 
one dimension,} J. of Diff. Equations, {\bf 137}(2), (1997), 251--272.



\bibitem{Hi} M. Hitrik, {\em Bounds on scattering poles in one dimension,}
Comm. Math. Phys. {\bf 208}(1999), 381--411.


\bibitem{FuRa} S. Fujiye and T. Ramond, {\em Matrice de scattering et
r\'esonances associ\'ees \`a une orbite h\'et\'erocline,} Ann.~IHP, Physiqie
Th\'eorique, {\bf 69}(1998), 31-82.


\bibitem{Kor} E. Korotyaev, {\em Inverse resonance scattering on the real line,}
Inverse Problems {\bf 21}(2005), 325--341.

\bibitem{ARPACK}
R.B. Lehoucq, D. Sorensen, and C. Yang,
 {\emph{ARPACK User's Guide: Solution of Large-Scale Eigenvalue Problems with
Implicitly Restarted Arnoldi Methods.}}\\
{\tt http://www.caam.rice.edu/software/ARPACK/UG/ug.html}
SIAM, 1998.

\bibitem{LL} L.D. Landau and E.M. Lifshitz, 
{\em Quantum Mechanics: Non-Relativistic Theory,} Third Edition, Elsevier, 1977.





\bibitem{Ned} L. Nedelec, {\em Asymptotics of resonances for 
matrix valued  Schr\"odinger operators,} {\tt math.SP/0509391}.


\bibitem{ThRa} T. Ramond, {\em Semiclassical study of quantum scattering
on the line,} Comm. Math. Phys. {\bf 177}(1996), 221--254.

\bibitem{Reg} T. Regge, {\em Analytic properties of the scattering matrix,}
Nuovo Cimento, {\bf 10}(1958), 671-679.


\bibitem{Si} B. Simon, 
{\em Resonances in one dimension and Fredholm 
determinants,} J. Funct. Anal. {\bf 178}(2000), 396--420.



\bibitem{TZ} S.H. Tang and M. Zworski, 
 {{\em Potential scattering on the 
real line,}}\\
{\tt http://math.berkeley.edu/$\sim$zworski/tz1.pdf}

\bibitem{Trefethen}
L.N. Trefethen, \emph{Spectral Methods in MATLAB,} SIAM, 2000.

\bibitem{Zw} M. Zworski, {\em Distribution of poles for scattering
on the real line,} J. Funct. Anal. {\bf 73}(1987), 277-296.






\end{thebibliography}
\end{document}